\documentclass[structabstract]{aa}  
\usepackage{graphicx}
\usepackage{textgreek}
\usepackage{hyperref}
\usepackage{txfonts}
\begin{document}

   \title{Concerning the occurrence of bow shocks around high-mass X-ray binaries \thanks{{\it Herschel} is an ESA space observatory with science instruments provided by European-led Principal Investigator consortia and with important participation from NASA.}}

   \author{M. Pri\v{s}egen
          }

   \institute{Department of Theoretical Physics and Astrophysics, Masaryk University, Kotl\'{a}\v{r}sk\'{a} 2, 611 37
              Brno, Czech Republic\\
              \email{michalprisegen@gmail.com}
             }

 
  \abstract
   {We investigate the occurrence of stellar bow shocks around high-mass X-ray binaries (HMXBs) in the Galaxy.}
   {We seek to conduct a survey of HMXBs in the mid-infrared to search for the presence of bow shocks around these objects.}
   {Telescopes operating in the mid-infrared, such as the \textit{Spitzer Space Telescope} or \textit{Wide-field Infrared Survey Explorer} (WISE), are potent 
    tools for searching for the stellar bow shocks. We used the available archival data from these telescopes to search for bow shock 
    candidates around the confirmed and candidate HMXBs in the Galaxy.}
   {We detected extended mid-infrared structures around several surveyed confirmed and candidate HMXBs. Two of these structures, associated with Vela X-1 and 4U 1907+09, are genuine bow shocks that have been studied previously. However, there are no new unambiguous bow shocks among the rest of the objects. The paucity of bow shocks around HMXBs suggests that the majority of these systems still reside within hot, low-density bubbles around their parent star clusters or associations. This also implies that the dynamical ejection of massive binaries is apparently less efficient than the ejections caused by the supernova explosions inside a binary.}
   {}


   \maketitle
%

\section{Introduction}

High-mass X-ray binaries (HMXBs) are systems consisting of a massive, early-type star (spectral class O or B) in the process of stellar evolution
and a compact object, neutron star, or  black hole, revolving around their common center of mass. The compact object accretes mass from the stellar wind of its companion or, if the companion star fills its 
Roche lobe, the matter flows directly onto the compact object (e.g., Lewin et al. \cite{lewin_97}; Lewin \& van
der Klis \cite{lewin_vdklis}; Liu et al. \cite{liu_2008}). 

Owing to the past supernova explosion of one of its components, possibly combined with close dynamical encounters of the system with other stars in its parent cluster as well, a significant
portion of HMXBs possess high (runaway) space velocities (e.g., van Oijen, \cite{oijen_1989}; Kaper et al., \cite{kaper_2004}). Furthermore, their galactic height distribution also implies their runaway nature (van Oijen \cite{oijen_1989}). 

Traditionally, runaway stars can be revealed via measurements of their proper motions, yielding
their tangential velocities. However, this method requires the distance to the object to
be determined and the proper motion to be measured with sufficient precision. Determination of the radial velocities via spectroscopic measurements is also a viable method. Nevertheless, these methods are only applicable to a portion of relatively close HMXBs, having sufficiently large proper motion to be reliably measured and/or bright optical counterpart for the spectroscopic studies. With the advent of modern high-energy missions, there has been a sharp increase in the number of confirmed and candidate HMXB systems that are detected at considerable distances, are often highly absorbed, and only have weak infrared counterparts. As they are hard objects to study with the above methods, the kinematics of these objects is poorly known.

 Another means of searching for runaway star candidates is through the detection of bow shocks associated with these objects (e.g., van Buren \& McCray \cite{vanburen_1988}; van Buren et al. \cite{vanburen_1995};  Gvaramadze \& Bomans \cite{gvaramadze_2008}; Gvaramadze et al. \cite{gvaramadze_2011a}; Peri et al. \cite{peri_2012}).  This method does not rely on the knowledge of the star's proper motion, precise distance, or the characteristics of the local standard of rest (LSR) of the studied system. Stellar bow shocks are also parsec-scale structures, manifesting strongly in the mid-infrared, so they can be detected at considerable distances due to their size and low levels of extinction at these wavelengths  (e.g., Gvaramadze et al. \cite{gvaramadze_2010}, Gvaramadze et al. \cite{gvaramadze_2011smc}). Thus, this method could be especially potent when studying distant HMXBs, as most of them do not have their proper motions and distances determined or they are measured with low significance.

The peculiar radial velocity of HMXBs, their Galactic height distribution, and low OB association memberships provides a strong evidence that HMXBs are a subclass of high-velocity stars (van Oijen \cite{oijen_1989}). This conclusion was verified by the peculiar tangential velocity measurements by Chevalier \& Ilovaisky (\cite{chevalier_1998}) and van den Heuvel et al. (\cite{heuvel_2000}). Both studies also note a disparity between the peculiar tangential velocities of OB-supergiant X-ray binaries and Be/X-ray binaries; the former group has a mean peculiar tangential velocity $<v_{\mathrm{pec,tan}}> \,=42 \pm 14 \, \mathrm{km\, s^{-1}}$, while the latter only has a modest $<v_{\mathrm{pec,tan}}> \,=15 \pm 6 \, \mathrm{km\, s^{-1}}$ (van den Heuvel et al. \cite{heuvel_2000}). Despite the considerable difference in the peculiar velocity between the two subgroups, the faster members of the Be/X-ray binary subgroup should have sufficient peculiar velocity to produce an observable bow shock (Meyer et al. \cite{meyer_2017}).  

The first detection of a bow shock associated with a HMXB was reported by Kaper et al. (\cite{kaper_1997}), who discovered a faint arcuate structure around a well-known HMXB \object{Vela X-1} in a H\textalpha \thinspace image obtained with the 1.54 m Danish telescope at the European Southern Observatory.
The first bow shock survey of HMXBs was conducted by Huthoff \& Kaper (\cite{huthoff_2002}), choosing a sample of 11 HMXBs for which the conditions for bow shock formation
seemed favorable: a large peculiar velocity and an OB companion with a strong stellar wind and a high luminosity. Using high-resolution \textit{IRAS} maps, Huthoff \& Kaper detected an infrared counterpart to the already known bow shock generated by \object{Vela X-1}.
Another survey was conducted by Gvaramadze et al. (\cite{gvaramadze_2011b}) on the same sample of HMXB systems. They utilized the data from the \textit{Spitzer Space Telescope}, which were of higher quality but they covered only 5 of 11 HMXB from the original sample.
Nevertheless, these authors were able to discover a new bow shock around \object{4U 1907+09} and provided a more detailed view of Vela X-1 bow shock as well.

We are currently aware of more than a hundred HMXBs or HMXB candidates in the Galaxy and their numbers are increasing (Liu et al. \cite{liu_2008}). Together with the advent of space infrared missions such as \textit{Spitzer Space Telescope} (Werner et al. \cite{werner_2004}) and \textit{Wide-field Infrared Survey Explorer} (WISE; Wright et al. \cite{wright_2010}), this
presents an opportunity to have a more detailed look at the bow shock occurence around HMXBs. In this paper, we have undertaken a  search for bow shocks around the Galactic HMXBs from the fourth edition of the Catalog of High Mass X-ray binaries in the Galaxy (Liu et al. \cite{liu_2008}) and around systems listed in Walter et al. (\cite{walter_2015}), which are not listed in Liu et al. (\cite{liu_2008}).

The paper is organized as follows. The rationale, observations, and general data processing are described in Sect. 2. A detailed description of the selected targets and supplementary astrometry is given, respectively, in Sect. 3 and Sect. 4. We discuss the nature of the emission around the studied objects in Sect. 5. Finally, the summary and conclusions are provided in Sect. 6.


\section{Rationale and methods}

A number of past successful studies of bow shocks conducted in the mid-infrared (e.g., Peri et al. \cite{peri_2012}, \cite{peri_2015}, Kobulnicky et al. \cite{kobulnicky_2016} and the references therein) and extensive computational work (e.g., Meyer et al. \cite{meyer_2016}, 
\cite{meyer_2017}) suggest that this part of the electromagnetic spectrum is the most suited for bow shock searches. 

To search for bow shocks around HMXBs, we used data obtained by Multiband Imaging Photometer for Spitzer (MIPS; Rieke et al. \cite{rieke_2004}) extracted from the \textit{Spitzer Space Telescope} archive and data obtained from WISE Image Service. The available \textit{Spitzer} data cover the fields containing less than a half of catalogued HMXBs, the data from WISE cover all surveyed HMXB systems. To obtain science-grade images, the \textit{Spitzer} data was processed by Mosaicking and Point source Extraction (MOPEX; Makovoz \& Khan \cite{makovoz_2005}). Visual inspection of \textit{Spitzer} and WISE images was used to search for the extended emission possibly associated with the HMXBs. To search for the possible counterparts of the detected emission, we used the SuperCOSMOS H-alpha Survey(SHS; Parker et al. \cite{parker_2005}), data obtained by the Photodetector Array Camera and Spectrometer (PACS; Poglitsch et al. \cite{poglitsch_2010}), and the Spectral and Photometric Imaging Receiver (SPIRE; Griffin et al. \cite{griffin_2010}) on board the \textit{Herschel Space Observatory} (Pilbratt et al. \cite{pilbratt_2010}),  and data obtained by the InfraRed Array Camera (IRAC; Fazio et al. \cite{fazio_2004}) on board the \textit{Spitzer Space Telescope}.

\section{Remarks on individual objects}
Owing to the short lifetimes of HMXBs, these objects cannot travel far from the Galactic plane where they formed. This means that they are often projected onto a region with complex diffuse infrared emission and it is often hard to conclude whether the emission is actually physically related to the system or is of a foreground/background origin. However, a physical relation is likely if the diffuse emission is centered at the system or exhibits some degree of symmetry wherein the HMXB lies on or near the axis of symmetry. Below, we briefly describe the systems with such infrared emission. 

\subsection{$\gamma$ Cas}
\object{$\gamma$ Cas} has attracted a lot of attention since its discovery as the first of what became known as Be stars (Secchi \cite{secchi_1866}). Despite considerable interest, the origin of X-ray emission from the object is still highly debated. The proposed scenarios include  the accretion onto a neutron star or a white dwarf and a magnetic star-disk interaction (see review by Smith et al. \cite{smith_2016} and the references therein). The nature of this system is so peculiar that it has become a prototype of a separate class of X-ray emitters knowns as `$\gamma$ Cas analogs'. While the recent study by Postnov et al. (\cite{postnov_2017}) reconciles the peculiar X-ray properties of \object{$\gamma$ Cas} by invoking a fast spinning neutron star as a companion, their results are disputed (Smith et al. \cite{smith_2017}). Because of its atypical properties and the disputed nature, we  consider \object{$\gamma$ Cas} as a HMXB candidate only.

\object{$\gamma$ Cas} is a relatively nearby object. Both Perryman et al. (\cite{perryman_1997}) and Coleiro \& Chaty (\cite{coleiro_2013}) put it at a similar distance of 0.19 and 0.17 kpc, respectively. Megier et al. (\cite{megier_2009}) estimated a lower distance of 0.12 kpc. We adopt a distance of 0.15 kpc as a compromise for the subsequent analysis.

\begin{figure*}
\sidecaption
 \includegraphics[width=12cm]{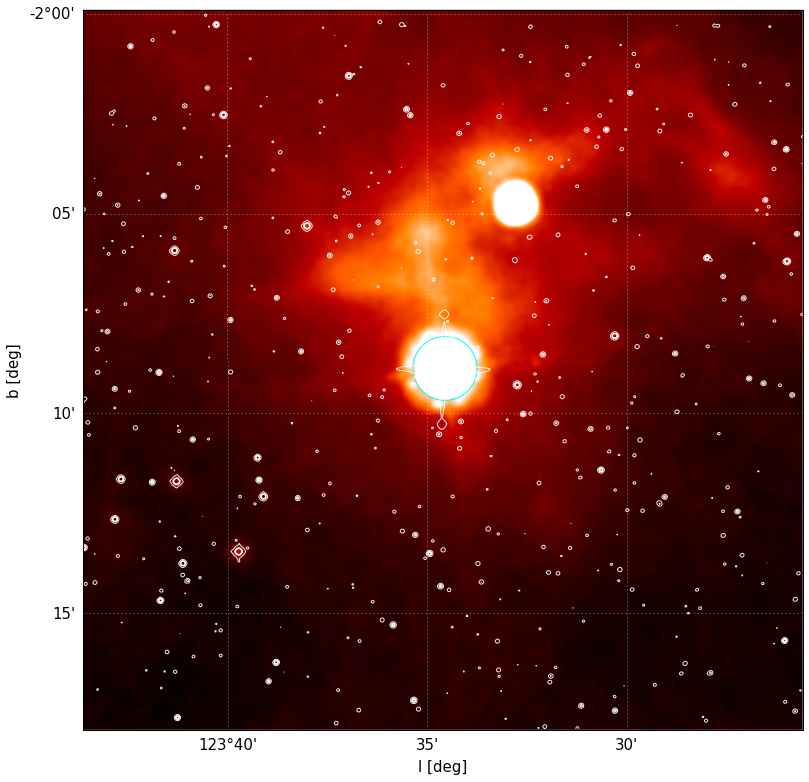}
  \caption{WISE 22 $\mathrm{\mu}$m image of the field around \object{$\mathrm{\gamma}$ Cas} with 2MASS Ks contours (sensitive to starlight) overlaid on top. The system position is indicated by the cyan circle. }
  \label{gamma_Cas_nebula}
\end{figure*}

WISE 22 $\mathrm{\mu}$m image (Fig.~\ref{gamma_Cas_nebula}) reveals a faint arc-like nebula to the north of the system. Unfortunately, the system is not covered by the Herschel data and Spitzer data are limited to the first two IRAC channels, taken in the post-cryo mode, which do not show any discernable diffuse emission. The system is slightly offset from the axis of the symmetry of the arc. The system is also embedded in a fainter 22 $\mathrm{\mu}$m diffuse emission.

\subsection{EXO 051910+3737.7 (V420 Aur)}
The X-ray source was discovered by Uhuru satellite (Forman et al. \cite{forman_1978}) and associated with a Be star \object{V420 Aur} by Polcaro et al. (\cite{polcaro_1984}). No X-ray pulses have been detected from the system (Liu et al. \cite{liu_2008}).

The system does not have a significant Hipparcos parallax, Chevalier \& Ilovaisky (\cite{chevalier_1998}) estimated the distance to be approximately 1.05 kpc using a typical luminosity of a B0 subgiant. Astraatmadja \& Bailer-Jones (\cite{Astraatmadja_2016}) estimated the distance to be 1.5 $\pm$ 0.5 kpc utilizing Tycho-Gaia astrometric solution (TGAS; Lindegren et al. \cite{lindegren_2016}) parallaxes and Milky Way prior.  A more recent catalog by Bailer-Jones et al. (\cite{bailer-jones_2018}), employing the second Gaia data release (hereafter GDR2, Gaia collaboration \cite{gaia_gdr2_summary}), puts the system at $\mathrm{1.29_{-0.09}^{+0.11}}$ kpc.

\begin{figure*}
\sidecaption
\includegraphics[width=12cm]{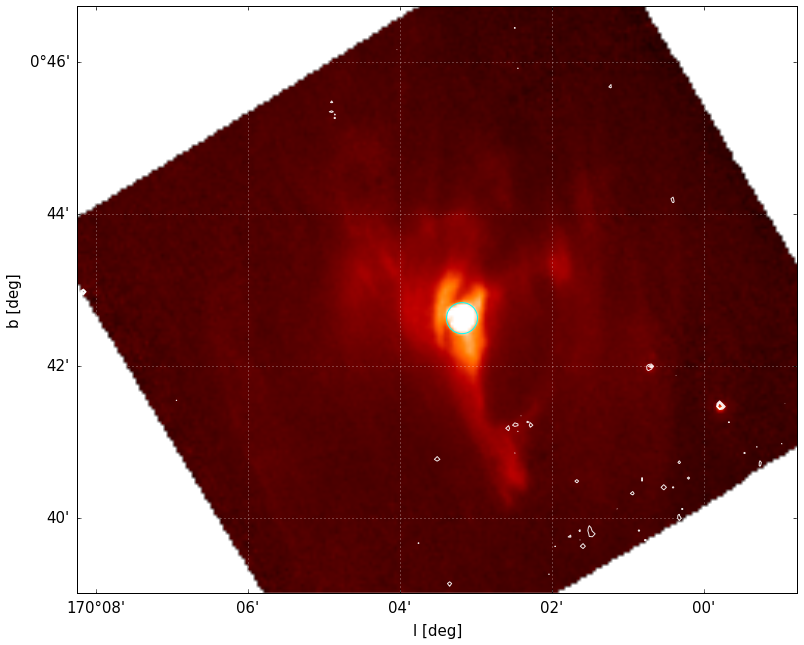}
\caption{\textit{Spitzer} MIPS 24 $\mathrm{\mu}$m image of the field around \object{EXO 051910+3737.7} (\object{V420 Aur}) with \textit{Spitzer} IRAC 3.6 $\mathrm{\mu}$m contours (mainly sensitive to starlight) overlaid on top.  The system position is indicated by the cyan circle.}
\label{V420_Aur_nebula}
\end{figure*}

The system appears to be a center of an irregular wispy infrared nebula (Fig.~\ref{V420_Aur_nebula}), readily apparent on WISE 12 $\mathrm{\mu}$m, 22 $\mathrm{\mu}$m, and MIPS 24 $\mathrm{\mu}$m images. The central, brighter part of the nebula, is divided into two filaments, aligned in the Galactic north-south direction: one apparently connected with the system and one slightly ahead of it. The nebula also possesses four fainter arms; the most dominant arm extends toward the Galactic south and the fainter three point to the Galactic north, northwest, and northeast. 

\subsection{IGR J11435-6109}
\object{IGR J11435-6109} is a HMXB discovered by Grebenev et al. (\cite{grebenev_2004}) via INTEGRAL mission. The system is an X-ray pulsar with a pulse period of about 162 s and orbital period of 52.5~d (in't Zand \& Heise \cite{int_zand_2004}, Corbet \& Remillard \cite{corbet_2005b}). The optical counterpart was determined by Tomsick et al. (\cite{tomsick_2007}) and confirmed by Negueruela et al. (\cite{negueruela_2007}). Coleiro et al. (\cite{coleiro_2013b}) refined the spectral classification to B0.5Ve, confirming the system as a Be/X-ray binary. 

The system is a distant X-ray binary ($>$ 6 kpc, Negueruela et al. \cite{negueruela_2007}). Masetti et al. (\cite{masetti_2009}) estimated the distance to be about 8.6 kpc. Coleiro \& Chaty (\cite{coleiro_2013}) derived a slightly higher distance, d = 9.8 $\pm$ 0.86 kpc. Bailer-Jones et al. (\cite{bailer-jones_2018}) confirmed the distant nature of this HMXB, placing it at $8.59_{-1.81}^{+2.54}$ kpc.

\begin{figure*}
\sidecaption
\includegraphics[width=12cm]{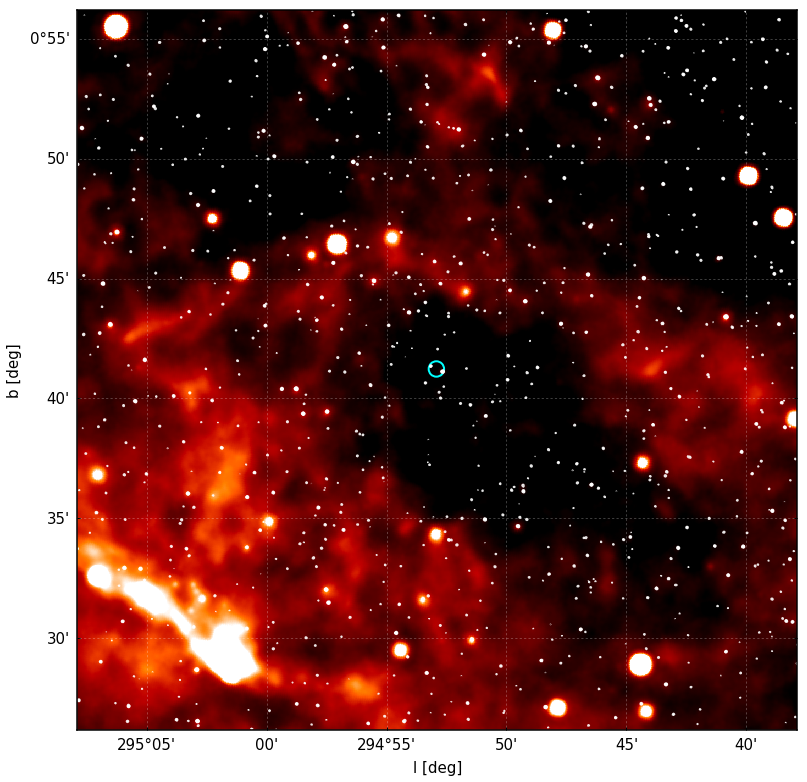}
\caption{WISE 22 $\mathrm{\mu}$m image of the field around \object{IGR J11435-6109} with 2MASS K band contours (sensitive to starlight) overlaid on top.  The system position is indicated by the cyan circle.}
\label{IGR_J11435-6109_nebula}
\end{figure*}

Fig.~\ref{IGR_J11435-6109_nebula} shows the enviroment around the system. \object{IGR J11435-6109} appears to be projected onto the eastern part of a large cavity. 

\subsection{GX 301-02}
\object{GX 301-2} is an obscured HMXB system that is very bright in X-rays thanks to the slow and very dense stellar wind of its massive, hyper giant companion (Kaper et al. \cite{kaper_2006}). The system contains a pulsar with a period of around 700 s in a 41.5 d, highly eccentric orbit (Sato et al. \cite{sato_1986}). 

The distance to \object{GX 301-2} was estimated by Kaper et al. (\cite{kaper_2006}) to be in a range of 3-4 kpc. Coleiro \& Chaty (\cite{coleiro_2013}) estimated the distance to be 3.1 $\pm$ 0.64 kpc via a spectral energy distribution procedure. The system is covered in TGAS, however, the measured parallax is not significant (0.34$\pm$0.75), so the distance estimates by Astraatmadja \& Bailer-Jones (\cite{Astraatmadja_2016}) vary significantly based on the prior used. The distance estimate was refined after GDR2; Bailer-Jones et al. (\cite{bailer-jones_2018}) estimated the distance to the system d = $3.53_{-0.40}^{+0.52}$ kpc, confirming the previous estimates.

\begin{figure*}
\sidecaption
\includegraphics[width=12cm]{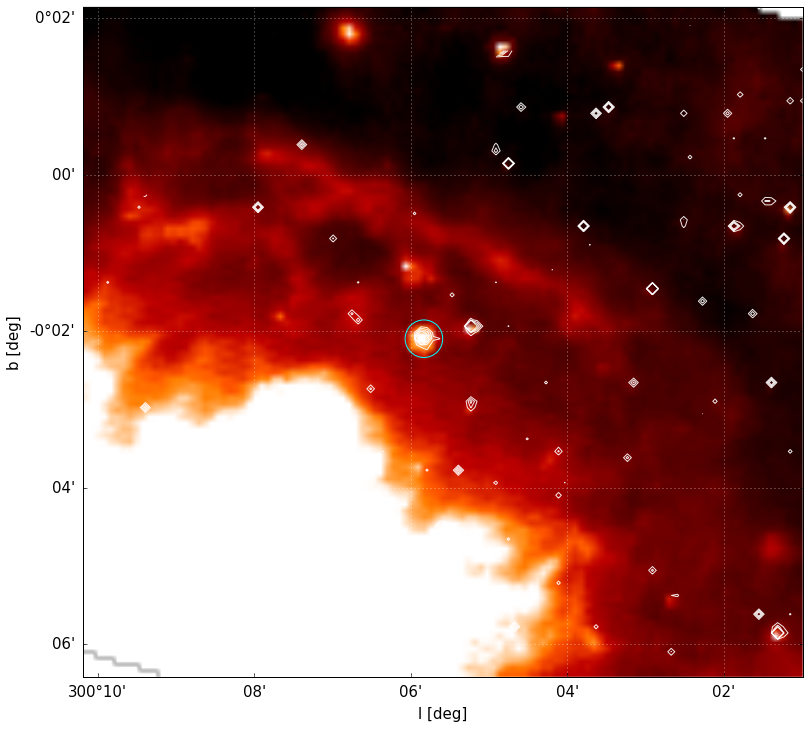}
\caption{\textit{Spitzer} MIPS 24 $\mathrm{\mu}$m image of the field around \object{GX 301-2} (\object{BP Cru}) with \textit{Spitzer} IRAC 3.6 $\mathrm{\mu}$m contours (mainly sensitive to starlight) overlaid on top.  The system position is indicated by the cyan circle.}
\label{GX_301-2_nebula}
\end{figure*}

The system lies in a region of very bright and complex infrared emission (see Fig.~\ref{GX_301-2_nebula}). This system is embraced by a region of bright 24 $\mathrm{\mu}$m emission to the Galactic south and a bright ridge of extended emission to the north. The PACS and SPIRE data suggest that the system could be enclosed in a partial far-infrared bubble, open at one side.

\subsection{1H 1255-567}
Listed as a HMXB candidate in Liu et al. (\cite{liu_2008}), but as a Be star in Simbad (Levenhagen \& Leister \cite{levenhagen_leister_2006}), the system is a part of a visual double with \object{$\mu^{1}$ Cru} (Hoffleit \& Jaschek \cite{hoffleit_1982}).

\object{1H 1255-567} appears to be a close system. Chevalier \& Ilovaisky estimated its distance to be about 111 $\pm$ 8 pc. Chen et al. (\cite{chen_2012}) listed a slightly larger value of 125 $\pm$ 5 pc. More recently, Bailer-Jones et al. (\cite{bailer-jones_2018}) estimated the distance to be $112_{-3}^{+2}$ pc.

\begin{figure*}
\sidecaption
\includegraphics[width=12cm]{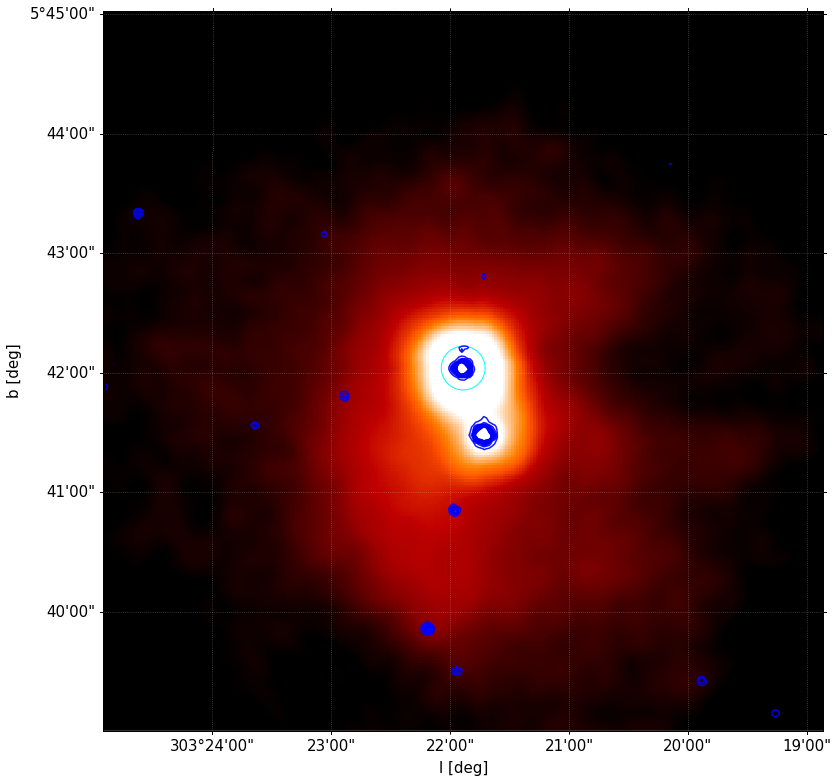}
\caption{WISE 22 $\mathrm{\mu}$m image of the field around \object{1H 1255-567} (\object{$\mathrm{\mu^{2}}$ Cru}) with 2MASS Ks contours (sensitive to starlight) overlaid on top. The system position is indicated by the cyan circle.}
\label{mu2_Cru_nebula}
\end{figure*}

Fig.~\ref{mu2_Cru_nebula} shows a WISE 22 $\mathrm{\mu}$m image of the visual double with both members visible in far-infrared emission, enshrouded in a fine extended nebula. The nebula bears a complex shape and has a system of fine filaments and wisps, emanating from \object{$\mathrm{\mu^{1}}$ Cru} (the prominent unmarked source to the Galactic south of \object{1H 1255-567}) and \object{1H 1255-567} as well.

\subsection{GX 304-01}
The HMXB \object{GX 304-01} (\object{4U 1258-61}) was discovered by balloon X-ray observations in 1970 (McClintock et al. \cite{mcclintock_1971}). The optical counterpart of the X-ray source was identified by Mason et al. (\cite{mason_1978}), making \object{V850 Cen} the most likely counterpart. A subsequent spectral analysis by Parkes et al. (\cite{parkes_1980}) showed the presence of a strong double-peaked emission in H\textalpha \thinspace and deep sharp absorption lines indicating a Be star. The most recent measurements give a more precise spectral classification that indicates the secondary is a B0.7 Ve star (Ziolkowski \cite{ziolkowski_2002}).  Using the data from Vela satellite, Priedhorsky \& Terrell (\cite{priedhorsky_1983}) found a periodicity in the X-ray flux of 133 d and interpreted this as being the orbital period of a neutron star in an orbit around the Be star, the variability in the X-ray flux being caused by enhanced accretion onto the neutron star as it passes through the periastron. This establishes \object{GX 304-01} as a HMXB with a Be star companion (i.e., a Be/X-ray binary).

 Parkes et al. (\cite{parkes_1980}) estimated the distance of the system to be 2.4  $\pm$ 0.5 kpc. Coleiro \& Chaty (\cite{coleiro_2013}) provided an estimate of the distance of 1.3 kpc. Using GDR2, Bailer-Jones et al. (\cite{bailer-jones_2018}) derived a larger distance of $2.01_{-0.13}^{+0.15}$ kpc.

\begin{figure*}
\sidecaption
\includegraphics[width=12cm]{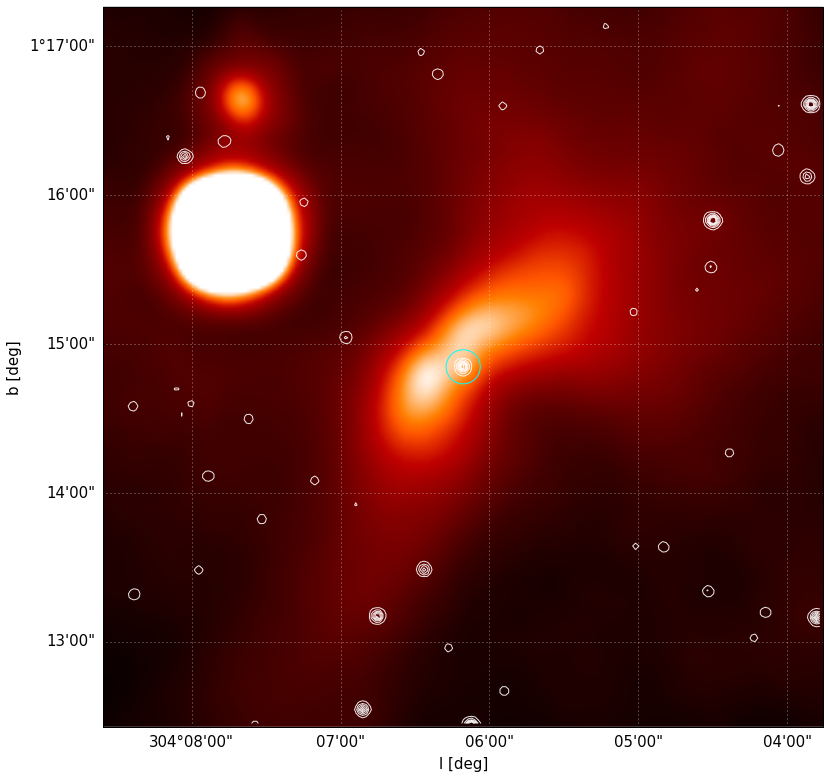}
\caption{WISE 22 $\mathrm{\mu}$m image of the field around \object{GX 304-01} (\object{V850 Cen}) with 2MASS Ks contours (sensitive to starlight) overlaid on top. The system position is denoted by the cyan circle. }
\label{GX_304-1_nebula}
\end{figure*}

The 22 $\mathrm{\mu}$m WISE image of the system (see Fig.~\ref{GX_304-1_nebula}) reveals a fine arcuate nebula that has a noticable dimple at the apex of the arc, close to the HMXB. Unfortunately, the system is not covered by the \textit{Spitzer} data. The arc is also visible in PACS 70 $\mathrm{\mu}$m image; no H$\mathrm{\alpha}$ counterpart was detected. The arc exhibits a clear symmetric morphology without any discernable clumps or finer structure. The infrared counterpart of \object{GX 304-01} lies along the axis of symmetry of the arc, making it  likely that this emission is associated with the HMXB system. Because of the apex dimple, limited resolution of WISE  22 $\mathrm{\mu}$m channel, and close proximity of the nebula to the system, the HMXB seems to be immersed in the nebula.

\subsection{AX J1639.0-4642}
The   X-ray  source \object{AX J1639.0-4642}  was  discovered   by   the   Advanced   Satellite   for Cosmology  and Astrophysics (ASCA) observatory (Sugizaki et al. \cite{sugizaki_2001}). The system is likely a supergiant HMXB.  It is highly absorbed X-ray pulsar with a spin period of 911 s (Bodaghee et al. \cite{bodaghee_2006}).  The infrared counterpart identification has been problematic. \object{2MASS J16390535-4642137} was proposed as a possible counterpart (Chaty et al. \cite{chaty_2008}). Further refinement of the system position by Bodaghee et al. (\cite{bodaghee_2012a}) suggested another close, faint, and blended star as the most probable counterpart. The precise orbital period of the system is disputed, but is most likely around 4.2 d (e.g., Corbet et al. \cite{corbet_2010}, Corbet \& Krimm \cite{corbet_2013}). These values of pulse and orbital period place the system among the other wind-fed HMXBs in Corbet's diagram of pulse versus orbital period (Corbet \cite{corbet_1986}).

The faintness of the infrared counterpart hampers the distance determination. Assuming a luminosity typical for an accretion-driven X-ray pulsar, Bodaghee et al. (\cite{bodaghee_2006}) estimated the distance of the source to be approximately 10 kpc.

\begin{figure*}
\sidecaption
\includegraphics[width=12cm]{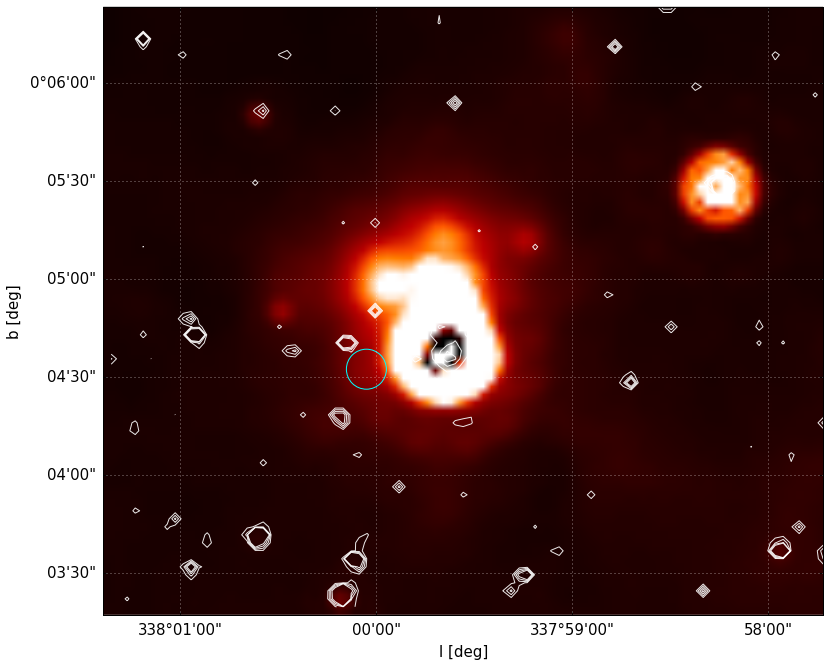}
\caption{\textit{Spitzer} MIPS 24 $\mathrm{\mu}$m image of the field around \object{AX J1639.0-4642} with \textit{Spitzer} IRAC 3.6 $\mathrm{\mu}$m contours (mainly sensitive to starlight) overlaid on top.  The system position is denoted by the cyan circle.}
\label{AX_J1639_nebula}
\end{figure*}

Fig.~\ref{AX_J1639_nebula} shows the field around the system. The HMXB is adjacent to a bright 24 $\mathrm{\mu}$m nebula, centered on the group of faint infrared stars. The HMXB is immersed into the fainter outskirts of this nebula. The nebula also dominates the field at IRAC and Herschel wavelengths.

\subsection{4U 1700-37}
\object{4U 1700-37} is an eclipsing HMXB with a very massive companion of O6Iafcp spectral class (Jones et al. \cite{jones_1973}, Sota et al. \cite{sota_2014}). The nature of the compact object is unclear.  The absence of pulsation suggests the presence of a low-mass black hole, but a neutron star seems more likely (e.g., Reynolds et al. \cite{reynolds_1999}, Boroson et al.\cite{boroson_2003}).

Ankay et al. (\cite{ankay_2001}) estimated the distance to be about 1.9 kpc. Megier et al. (\cite{megier_2009}) and  Coleiro \& Chaty (\cite{coleiro_2013}) adopted 2.12 $\pm$ 0.34 kpc and 1.8 $\pm$ 0.15 kpc, respectively. Depending on the adopted prior, Astraatmadja \& Bailer-Jones (\cite{Astraatmadja_2016}) derived vastly different distances from the TGAS data (approximately from 0.7 to 3.3 kpc), which also suffer from large uncertainties. Utilizing GDR2, Bailer-Jones et al. (\cite{bailer-jones_2018}) derived a more precise distance estimate of $1.75_{-0.19}^{+0.24}$ kpc.

\begin{figure*}
\sidecaption
\includegraphics[width=12cm]{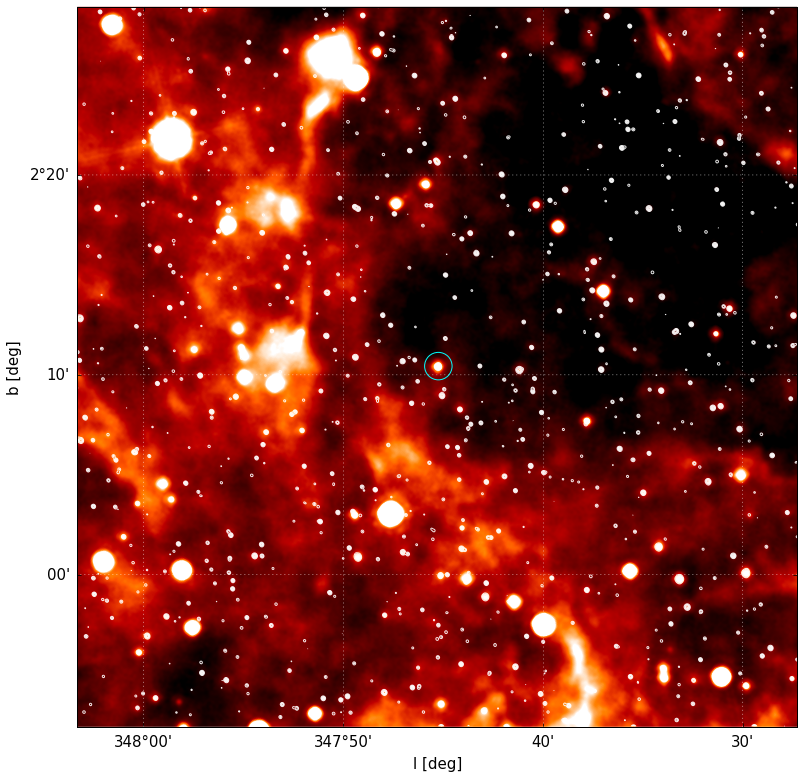}
\caption{WISE 22 $\mathrm{\mu}$m image of the field around \object{4U 1700-37} (\object{V884 Sco}) with WISE 3.4 $\mathrm{\mu}$m contours (mainly sensitive to starlight) overlaid on top. The system position is denoted by the cyan circle.}
\label{4U_1700-37_nebula}
\end{figure*}

Fig.~\ref{4U_1700-37_nebula} shows that the system lies on the axis of symmetry of an arcuate cavity that is excavated into a nearby bright  cloud of 24 $\mathrm{\mu}$m. 

\subsection{EXO 1722-363}
\object{EXO 1722-363} was discovered in the Galactic ridge by EXOSAT observations in 1984 (Warwick et al. \cite{warwick_1988}). Analysis of Ginga observations by Tawara et al. (\cite{tawara_1989}) suggested the presence of a dense envelope of circumstellar matter around the system. Takeuchi et al. (\cite{takeuchi_1990}) placed a 9 d lower limit for the orbital period, which was later refined by Corbet et al. (\cite{corbet_2005}) and Thompson et al. (\cite{thompson_2007}) to be 9.74 d. Using INTEGRAL observations, Zurita Heras et al. (\cite{zuritaheras_2006}) found a probable infrared counterpart to the system, \object{2MASS J17251139-3616575}, 1$^{\prime\prime}$  away from the best position of the source as given by INTEGRAL. Subsequent observations of the counterpart by Rahoui et al. (\cite{rahoui_2008}) and Mason et al. (\cite{mason_2009})  supported the earlier assumptions that the system is a HMXB containing an early supergiant B star, which produces strong stellar wind that fuels the accretion onto the neutron star.

The absence of the visual counterpart and the location of the system in the Galactic ridge region make the distance determination difficult. Zurita Heras et al. (\cite{zuritaheras_2006}), considering a typical luminosity of an active accretion pulsar, estimated the distance to the system to be 7 kpc. Thompson et al. (\cite{thompson_2007}) placed the system between 5.3 and 8.7 kpc. Using the broadband data, Rahoui et al. (\cite{rahoui_2008}) derived a slightly smaller distance to the system of 6.1 kpc. Distance determinations were also made by Mason et al. (\cite{mason_2009}), later refined by Mason et al. (\cite{mason_2010}), placing the system to be 7.1 $\leq d \leq$ 7.9 kpc. 

\begin{figure*}
\sidecaption
\includegraphics[width=12cm]{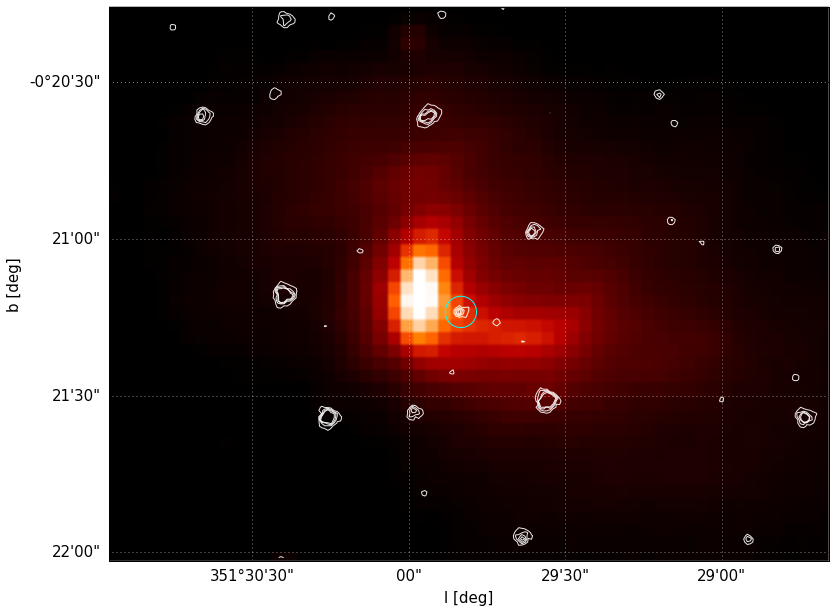}
\caption{\textit{Spitzer} MIPS 24 $\mathrm{\mu}$m image of the field around \object{EXO 1722-363} with \textit{Spitzer} IRAC 3.6 $\mathrm{\mu}$m contours (mainly sensitive to starlight) overlaid on top.  The system position is indicated by the cyan circle.}
\label{EXO_1722-363_nebula}
\end{figure*}

The nebula around \object{EXO 1722-363} is visible in both the \textit{Spitzer} MIPS 24 $\mathrm{\mu}$m (see Fig.~\ref{EXO_1722-363_nebula}) and 22 $\mathrm{\mu}$m WISE images. The arc is also visible in the PACS 70 $\mathrm{\mu}$m image and a gleam of emission possibly associated with the arc is also visible in the IRAC 8 $\mathrm{\mu}$m image. No H$\mathrm{\alpha}$ counterpart was detected. The \textit{Spitzer} MIPS 24 $\mathrm{\mu}$m image reveals an irregular, curved morphology, where the part of the nebula to the Galactic east of the system appears to be brighter and more prominent. This seems to be a feature of the structure itself, as our inspection of this area in the different bands does not reveal any possible background or foreground sources that could be responsible for the enhanced emission. The infrared counterpart of \object{EXO 1722-363} appears to lie within the nebula. If we consider the outer boundary of the nebula tracing an arc, the system lies along the approximate symmetry axis of this arc. This, together with the lack of any notable brighter point sources within the nebula, suggests that the structure is likely associated with the HMXB. 

\subsection{XTE J1739-302}
\object{XTE J1739-302} (\object{IGR J17391-3021}) is a supergiant X-ray transient (SFXT) discovered with the proportional counter array on board the \textit{Rossi X-Ray Timing Explorer} (RXTE; Smith et al. \cite{smith_1998}). The system consists of a neutron star on an eccentric 51.47 d orbit (Drave et al. \cite{drave_2010}) around a O8 Iab(f) supergiant companion (Negueruela et al. \cite{negueruela_2006a},  Rahoui et al. \cite{rahoui_2008}).

The source lies in the direction of the Galactic center. Using spectroscopy and photometry of the counterpart, Negueruela et al. (\cite{negueruela_2006a}) estimated its distance to be 2.3 $\pm$ 0.6 kpc. The subsequent analysis by Rahoui et al. \cite{rahoui_2008} derived a slightly larger distance of about 2.7 kpc. The distance estimate from GDR2 seems to be less constraining. Bailer-Jones et al. (\cite{bailer-jones_2018}) provided an estimate for the distance of $5.32_{-2.12}^{+3.92}$ kpc. It is therefore questionable whether the HMXB is a foreground source, as suggested by the earlier studies, or is, potentially, a Galactic center object.

\begin{figure*}
\sidecaption
\includegraphics[width=12cm]{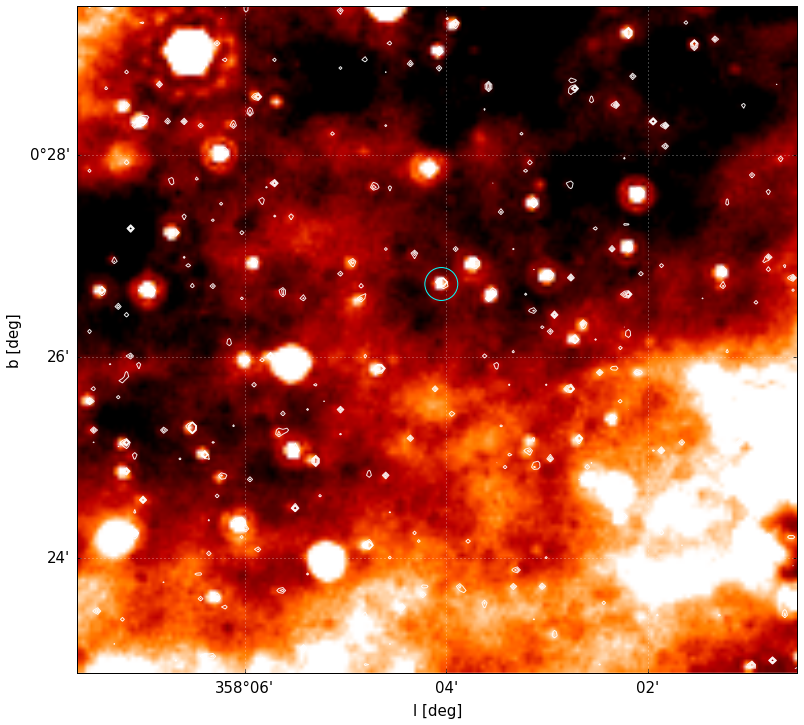}
\caption{\textit{Spitzer} MIPS 24 $\mathrm{\mu}$m image of the field around \object{XTE J1739-302} with \textit{Spitzer} IRAC 3.6 $\mathrm{\mu}$m contours (mainly sensitive to starlight) overlaid on top.  The system position is indicated by the cyan circle.}
\label{XTE_1739_nebula}
\end{figure*}

Fig.~\ref{XTE_1739_nebula} shows that the system lies in a region of complicated mid-infrared emission, typical for Galactic center region.  The system could be lying on the axis of symmetry of a faint arc of extended emission or a partial bubble.

\subsection{AX J1841.0-0536}
\object{AX J1841.0-0536} was discovered as a variable X-ray source by ASCA (Bamba et al. \cite{bamba_2001}) while undergoing two bright flares lasting approximately 1 hour each. The flaring activity of the system puts it into a SFXT class of HMXBs (e.g., Romano et al. \cite{romano_2011}).  Chandra observations by Halpern et al. (\cite{halpern_2004}) pinpointed the infrared counterpart. Negueruela et al. (\cite{negueruela_2006b}) provided constraints on the spectral type of the secondary, being of a luminous B0-1 type. This is further refined by the infrared spectroscopy done by Nespoli et al. (\cite{nespoli_2008}), who derived the spectral type B1Ib for the secondary. The  orbital period of the system of 6.45 d was reported by Gonz\'{a}lez-Gal\'{a}n (\cite{gonzales-galan_2015}).

The distance estimates of the system vary. Nespoli et. al (\cite{nespoli_2008}) provided a broad estimate of the distance of $3.2^{+2}_{-1.5}$ kpc. A more recent analysis by Coleiro \& Chaty (\cite{coleiro_2013}) put the system further away, at  7.8 $\pm$ 0.74 kpc . This is in agreement with the distance estimated from GDR2, d = $7.60_{-2.16}^{+3.06}$ kpc (Bailer-Jones et al. \cite{bailer-jones_2018}).

\begin{figure*}
\sidecaption
\includegraphics[width=12cm]{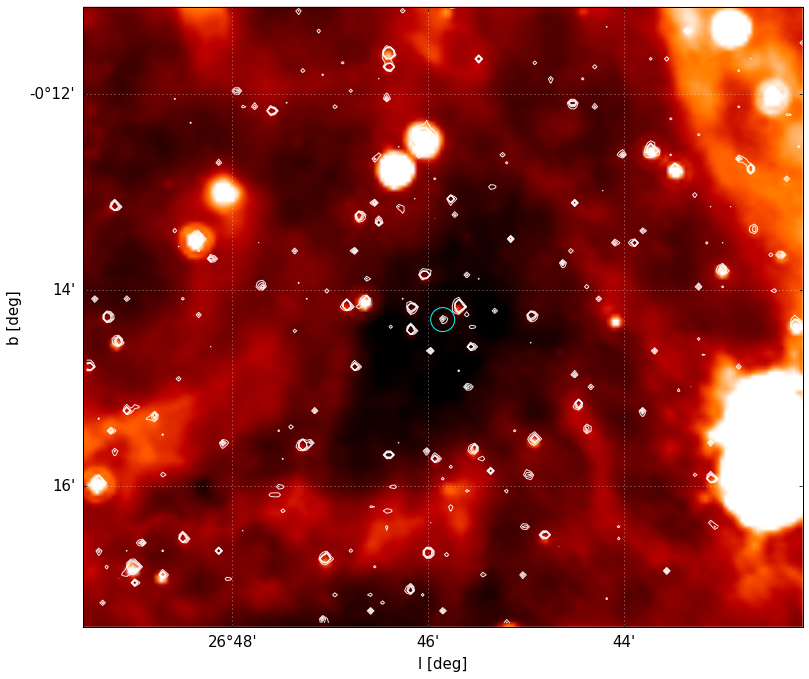}
\caption{\textit{Spitzer} MIPS 24 $\mathrm{\mu}$m image of the field around \object{AX J1841.0-0536} with \textit{Spitzer} IRAC 3.6 $\mathrm{\mu}$m contours (mainly sensitive to starlight) overlaid on top.  The system position is indicated by the cyan circle.}
\label{AX_J1841_nebula}
\end{figure*}

Fig.~\ref{AX_J1841_nebula} shows that the system lies in the center of the region of reduced 24 $\mathrm{\mu}$m emission. The cavity does not feature bright rims of mid-IR emission and the evidence of the reduced emission can be seen on the PACS and SPIRE wavebands as well.

\subsection{XTE J1855-026}
\object{XTE J1855-026} is a HMXB discovered by \textit{RXTE}. The system showed pulsations with a period of about 361 s and a flux modulation with a period of 6.1 d, which was interpreted as the orbital period of the system (Corbet et al. \cite{corbet_1999}). This was confirmed by Negueruela et al. (\cite{negueruela_2008}). The optical counterpart of the system was pinpointed by  Verrecchia et al. (\cite{verrecchia_2002}) and its spectral type was determined to be B0Iaep (Negueruela et al. \cite{negueruela_2008}).

Corbet et al. (\cite{corbet_1999}) proposed a distance of approximately 10 kpc. The analysis by Coleiro \& Chaty (\cite{coleiro_2013}) derived a consistent distance estimate of 10.8 $\pm$ 1.0 kpc. The distance given by Bailer-Jones et al. (\cite{bailer-jones_2018}) is slightly smaller, $8.14_{-1.79}^{+2.58}$ kpc.

\begin{figure*}
\sidecaption
\includegraphics[width=12cm]{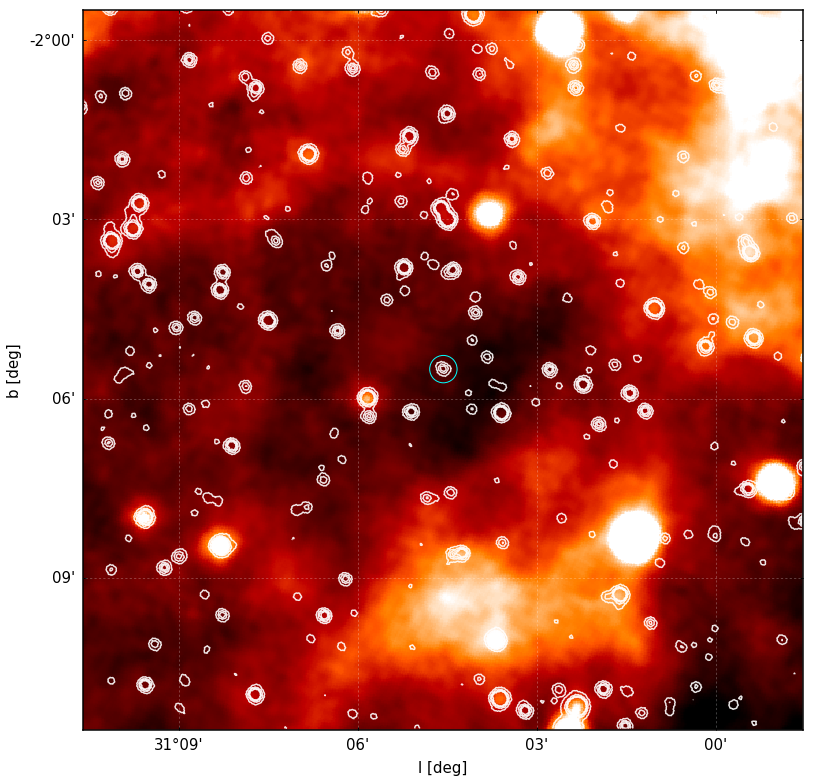}
\caption{WISE 22 $\mathrm{\mu}$m image of the field around \object{XTE J1855-026} with WISE 3.4 $\mathrm{\mu}$m contours (mainly sensitive to starlight) overlaid on top. The system position is indicated by the cyan circle.}
\label{XTE_J1855_nebula}
\end{figure*}

This system also appears to be associated with a region of reduced 24 $\mathrm{\mu}$m emission (Fig.~\ref{XTE_J1855_nebula}). The paucity of emission is not that well pronounced as is the case with the previous system, however, the region appears to be surrounded by a bright partial mid-infrared shell, most noticable to the Galactic south and northwest of the system.

\section{Kinematics of the studied systems}

The kinematics of the studied HMXBs is of significant interest, as it can provide constraints for several interesting system characteristics, such as the origin of the system or the properties of the progenitor binary. However, studying it has been difficult owing to considerable distances of these sources and insufficient accuracy of the past astrometric surveys. This resulted in the proper motions measured with a low significance, often varying significantly depending on the survey, making them hard to utilize. Only a few, very close sources did not suffer from these issues. The situation improved somewhat with the advent of the first Gaia data release (GDR1; Gaia Collaboration \cite{Gaia16}). While TGAS,  part of GDR1 (Lindegren et al. \cite{lindegren_2016}), contained only a small number of HMXBs, re-reductions of past catalogs using GDR1, such as UCAC5 (Zacharias et al. \cite{zacharias_2017}) and Hot Stuff for One Year (HSOY; Altmann et al. \cite{altmann_2017}) provided more precise proper motions for many more sources. However, the issue of discrepant proper motions depending on the catalog used still persisted. 

We used the parallaxes and proper motions from GDR2 (Gaia collaboration \cite{gaia_gdr2_summary}) to compute peculiar tangential velocities, which are especially interesting for systems associated with an arcuate structure in their vicinity. For our objects, we followed the approach outlined in Luri et al. (\cite{luri_2018}). The distances and tangential velocities were jointly estimated from the parallaxes and proper motions via Bayesian inferrence and the prior scale lengths are adopted from Bailer-Jones et al. (\cite{bailer-jones_2018}). To obtain the peculiar tangential velocities we adopted the Galactic constants $R_{0} = 8.2$ kpc, $\Theta_{0} = 238\, \mathrm{km}\, \mathrm{s}^{-1}$ and the solar peculiar motion $(U_{\odot}, V_{\odot}, W_{\odot}) = (10.0, 11.0, 7.0)\, \mathrm{km}\, \mathrm{s}^{-1}$ from Bland-Hawthorn \& Gerhard (\cite{bland-hawthorn_2016}). Table~\ref{HMXB_gaia} lists the computed peculiar velocity medians with the errors obtained from 68\% credibility intervals. The only exception was $\gamma$ Cas, as it is too bright for Gaia. In this case, we used the proper motion given in HSOY with an adopted distance of 0.15 kpc as discussed above. It must be noted that we computed the errors in the peculiar velocity for $\gamma$ Cas from the errors of the proper motion measurements only, so they should be considered as low-limit approximations.

Unfortunately, for the rest of the systems not listed in Table~\ref{HMXB_gaia} the choice of catalogs is limited. Because of the significant distance of these sources or faint optical/infrared counterparts, they are not listed in any proper motion catalogs or they exhibit no measurable proper motion.

\begin{table*}
\caption{Proper motions and derived peculiar tangential velocities for selected HMXBs. $\gamma$ Cas is separated from the rest of the systems owing    to its questionable nature and not having a solution in GDR2. In this case, the distance estimate is used instead of the parallax.}
\label{HMXB_gaia}
\centering
\begin{tabular}{c c c c c c}
\hline\hline
  ID & pmRA & pmDE  & parallax  & $v_{l, pec}$   & $v_{b, pec}$   \\
  & (mas/yr) & (mas/yr) & (mas) & ($\mathrm{km}\, \mathrm{s}^{-1}$) & ($\mathrm{km}\, \mathrm{s}^{-1}$) \\
  \hline
  EXO 051910+3737.7 & 1.44$\pm$0.12 & -4.12$\pm$0.07 & 0.753$\pm$0.057 & $15.5_{-2.0}^{+2.3}$ & $-0.4_{-1.0}^{+1.0}$ \\ 
  IGR J11435-6109 & -5.982$\pm$0.059 & 1.089$\pm$0.056 & 0.026$\pm$0.042 & $-10_{-63}^{+55}$ & $-13_{-5}^{+5}$ \\
  GX 301-02 & -5.303$\pm$0.051 & -2.166$\pm$0.049 & 0.253$\pm$0.035 & $34_{-23}^{+25}$ & $-45_{-8}^{+7}$ \\
  1H 1255-567 & -28.15$\pm$0.22 & -10.34$\pm$0.34 & 8.95$\pm$0.23 & $1.7_{-0.5}^{+0.4}$ & $1.9_{-0.2}^{+0.2}$ \\
  GX 304-01 & -4.235$\pm$0.043 & -0.316$\pm$0.043 & 0.470$\pm$0.033 & $24_{-5}^{+6}$ & $4.9_{-0.5}^{+0.5}$ \\
  4U 1700-37 & 2.220$\pm$0.087 & 4.954$\pm$0.073 & 0.0549$\pm$0.064 & $65_{-6}^{+7}$ & $17.3_{-1.4}^{+1.8}$ \\
  XTE J1739-302 & -0.89$\pm$0.22 & 3.49$\pm$0.17 & 0.12$\pm$0.16 & $161_{-80}^{+425}$ & $99_{-32}^{+48}$ \\
  AX J1841.0-0536 & -2.66$\pm$0.25 & -5.36$\pm$0.21 & -0.29$\pm$0.13 & $18_{-124}^{+122}$ & $4_{-8}^{+7}$ \\
  XTE J1855-026 & -2.605$\pm$0.063 & -6.788$\pm$0.056 & 0.039$\pm$0.044 & $-10_{-93}^{+83}$ & $-26_{-7}^{+6}$ \\
 \hline
 ID & pmRA  & pmDE  & distance  & $v_{l, pec}$  & $v_{b, pec}$  \\
 & (mas/yr) & (mas/yr) & (kpc) & ($\mathrm{km}\, \mathrm{s}^{-1}$) & ($\mathrm{km}\, \mathrm{s}^{-1}$) \\
 \hline
 $\gamma$ Cas & 24.950 $\pm$ 0.167 &  -3.890 $\pm$ 0.231  & 0.15 & 6.5 $\pm$ 0.1 & 5.0 $\pm$ 0.2\\
\hline
\end{tabular}
\end{table*}

\subsection{Astrometric flags in Gaia data release 2}
A number of the studied objects exhibit increased values of the astrometric flags, such as the astrometric excess noise and goodness-of-fit statistic, possibly indicating problems with the astrometric solution in GDR2. Issues with the astrometry may arise when dealing with the regions with large source densities, such as the Galactic plane and center regions, where HMXBs are predominantly situated. Another caveat is that the astrometric solution in GDR2 does not take the binarity into account, which may have an impact on the astrometry and its quality (Lindegren et al. \cite{lindegren_2018}). To investigate these effects, we queried GDR2 for stars of similar magnitude and color ($\mathrm{\Delta G = \pm 0.5}$ mag, $\mathrm{\Delta (BP-RP) = \pm 0.2}$ mag) within a 10 degrees radius cone centered on each source.  For each source we also constructed an equivalent query, extracting the sources from the opposite side of the sky to study the effects of crowding. Comparison of the studied HMXBs with the extracted stars from the source vicinity and the opposite side of the sky showed that, except for \object{AX J1841.0-0536}, none of the HMXBs are significant outliers, despite some sources having astrometric flag values indicating problems with the single-star astrometric solution. It is, however, unlikely that the measured parallaxes and proper motions are corrupted by the unmodeled orbital motion. The longest orbital period among our sources is approximately 133 d for \object{GX 304-01} (Priedhorsky \& Terrell \cite{priedhorsky_1983}). Even this orbital period is small compared to the 22-month observing time of GDR2. For such periods, it is expected that the orbital motion should average out and should not significantly impact the parallax and proper motion measurements (Jennings et al. \cite{jennings_2018}). Interestingly, \object{AX J1841.0-0536}, the system with elevated astrometric flags as compared to the extracted sources from the queries, only has an orbital period of 6.45 d  (Gonz\'{a}lez-Gal\'{a}n \cite{gonzales-galan_2015}).

\section{Discussion}
The infrared survey of the HMXB sample yielded a variety of extended emission that may suggest the influence of the HMXBs on the surrounding interstellar medium (ISM). This emission, its presence or absence, has  important implications not only for the particular system, but also for the HMXB population as a whole. 

\subsection{Extended emission}
The studied objects are projected against complex infrared structures, which are not only arcs bending outward from the HMXB, possibly indicating a bow shock, but also infrared filaments that appear to emanate from the system and emission cavities. 

\subsubsection{$\gamma$ Cas arc}

Fig.~\ref{gamma_Cas_nebula} shows the arcuate nebulosity in the vicinity of \object{$\gamma$ Cas}. While the infrared arcs around early-type stars are traditionally interpreted as stellar bow shocks, various phenomena can produce morphologies resembling bow shocks, including partial infrared bubbles and H II regions with density gradients. The infrared arcs may also be interpreted as dust waves or bow waves that may form around late-type OB main sequence stars, which may drive weaker stellar winds than expected (the weak wind problem; Ochsendorf et al. \cite{ochsendorf_2014a}, Ochsendorf et al. \cite{ochsendorf_2014b}). These structures can be attributed to the interaction of radiation pressure from the star with dust carried along by a photoevaporative flow of ionized gas from a dark cloud or inside an evolved interstellar bubble. The structures are very similar in morphology to genuine stellar bow shocks.  

The position of the arc is approximately consistent with the position of \object{GCRV 309 E}, a HII region, however, we were unable to confirm its presence on the image data in the INT Photometric H-Alpha Survey (IPHAS; Barentsen et al. \cite{barentsen_2014}) H$\alpha$ images. The absence of other infrared data makes it hard to establish the exact nature of the nebula. However, given the peculiar velocity of the source, its spectral type, a bow or a dust wave interpretation is more likely than a stellar bow shock. Adding to this, the arc approximately points to the the infrared pillars, associated with two clouds, \object{IRAS 00560+6037} and \object{IC 59}, which may be sources of a material flow toward the system. While the nature of the arc is debatable, it is, most likely, not a stellar bow shock.

\subsubsection{EXO 051910+3737.7 nebula}
The wispy nebula, fanning approximately in the Galactic north-south direction, may suggest a dusty outflow indicating a systemic mass loss. The nebula bears some resemblance to circumstellar nebulae observed around some B[e] and Be stars in the infrared images (Mayer et al. \cite{mayer_2016}). The literature on the enviroments around Be stars is sparse, but at least one Be star, \object{SX Aur}, also shows the presence of a mid-infrared nebula (Deschamps et al. \cite{deschamps_2015}), however, this nebula is more compact and shows a blob-like morphology instead of fine wispy jets.

There also exists a possibility that the dusty structure is not associated with the system and the system is just passing in its vicinity, heating the dust. However,  the position of the system in the central knot of the emission hints at the association. Moreover, the system has a significant peculiar velocity to the Galactic east. If the outflows emanate from the system, it would be expected to observe some bending of the structure toward the west, as the system plows through the surrounding ISM. This bending is apparent in the bright part of the nebula and also in the fainter south wisp. 

The nature of the easternmost rim of the nebula is puzzling. It is leading the star in the direction of its peculiar motion, so it cannot be related to the past mass-loss episode. It also exhibits the same degree of bending as the bright bar centered on the system. A bending toward the system due to its systemic peculiar motion would be expected for a stellar bow shock, however, the bar does not seem to be completely detached from the system, which is not expected. The bar may be a part of an infrared filament, locally heated by the system as it passes in its vicinity.

\subsubsection{IGR J11435-6109 cavity}
The system lies in a large infrared cavity that is apparent in WISE 12, 22 $\mathrm{\mu}$m images and in SPIRE wavebands. Coleiro et al. (\cite{coleiro_2013b}) discussed the cavity briefly and noted that it could be a result of the feedback of the system on the surrounding ISM. The physical relation of the system to the cavity is difficult to prove because of the inherent inaccuracy of the distance determination to both infrared cavities and distant HMXBs. Moreover, the system lies significantly off-center of the cavity. This could be reconciled if the system has a significant peculiar velocity pointing away from the cavity center. The system has proper motion data available but they are of a low quality. However, the peculiar motion component in the Galactic latitude is sufficiently well constrained to suggest that the system is moving to the Galactic south, which is at odds with its supposed origin near the center of the cavity. Considering the points above, it seems that the HMXB and the cavity it is projected against are unrelated.

\subsubsection{GX 301-02 nebula}
The system is surrounded by an extended infrared emission. Huthoff \& Kaper (\cite{huthoff_2002}) and Servillat et al. (\cite{servillat_2014}) discussed the possibility of a bow shock or a cavity associated with the system. The peculiar velocity obtained from the proper motion suggests that the system is moving predominantly to the Galactic southeast toward the bright extended emission, so the bright rim of infrared emission to the north of the system is unlikely to be a stellar bow shock. The surrounding emission is most likely related to the infrared bubble \object{[SPK2012] MWP1G300134-001035} to the south of the system. No bow shock is apparent despite the considerable peculiar velocity of the system.

\subsubsection{1H 1255-567 nebula}
The system is enshrouded in a fine wispy nebula, similar to the wispy nebula associated with \object{EXO 051910+3737.7}, but less prominent. Moreover, it is hard to determine if the emission is instead associated with \object{$\mathrm{\mu^{1}}$ Cru}.  The system exhibits no significant peculiar motion, yet the filaments seem to exhibit some curvature, as is evident in Fig.~\ref{mu2_Cru_nebula}. However, also the southern star, \object{$\mathrm{\mu^{1}}$ Cru}, does not exhibit any significant peculiar systemic motion.

\subsubsection{GX 304-01 arc}
The HMXB appears to be associated with a mid-infrared smooth arc pointing approximately to the Galactic northeast. The proper motion of the system from the GDR2 data yields a mildly runaway peculiar velocity primarily to the Galactic east, which deviates about 30 degrees from the approximate symmetry axis of the arc. The arcuate mid-infrared nebula upstream of the system could be interpreted as a stellar bow shock. However, there are several problems with this interpretation. The arc does not bear a classical bow shock shape and does not seem to be fully detached from the system. The most puzzling feature of the nebula is its apex dimple, pointing toward the system. Recently, Meyer et al. (\cite{meyer_2017}) conducted a series of simulations of stellar bow shocks of early-type stars in a magnetized medium. Their modeling suggests that in a magnetized ambient medium, the classic bow shock shape gets distorted and compressed, increasing its opening angle and becoming much blunter around the apex, as the stand-off distance decreases significantly. Interestingly, an apex dimple may form. This bow shock morphology change is especially prominent for stars having a modest space velocity, matching that of \object{GX 304-01}. This makes the bow shock interpretation appealing, however, the apparent attachment of the emission onto the system still remains an issue despite the expected decreased stand-off distance expected for such system. The mentioned misalignment between the peculiar velocity vector and the symmetry axis of the arc also poses a problem for the bow shock hypothesis, however, this can be reconciled by the presence of a large-scale flow in the ISM. An alternative origin of the emission might be due to the system encountering an infrared fillament along its way, locally heating and compressing it. An example of such system is \object{HD 49662}, as discussed in Kalas et al. (\cite{kalas_2002}). This system appears to be embedded in an infrared fillament, heating it locally. This produces a blister-like infrared emission, visible in WISE 12 $\mathrm{\mu}$m and 22 $\mathrm{\mu}$m images, bulging ahead of the system while the system is embedded in the diffuse emission. However, the peak of the diffuse emission is centered directly on the system, which is not the case for the HMXB. Also, the diffuse emission around \object{GX 304-01} does not bear a blister-like shape. Thus, we classify the emission around \object{GX 304-01} as ambiguous. 

\subsubsection{Blob near AX J1639.0-4642}
\object{AX J1639.0-4642} is adjacent to a blob of strong mid-infrared emission. The emission is very concentrated, however, its outskirts reach the HMXB, which is the most apparent in the longer IRAC wavebands. Several infrared point sources are present in the central part of the blob. This region is designated \object{IRAS 16353-4636} and is a site of star formation (Benaglia et al. \cite{benaglia_2010}). The point sources  within the blob make up a protostellar cluster. Benaglia et al. (\cite{benaglia_2010}) derived a distance of $\sim$ 8 kpc to the embedded protocluster. Owing to the inherent evolved nature of HMXBs, and because this protocluster is about 2 kpc closer than the HMXB, this cluster is not related to \object{AX J1639.0−4642}.

\subsubsection{4U 1700-37 cavity}
The system seems to be centered on an arcuate cavity protruding into a nearby mid-infrared cloud to the Galactic southeast. Toal{\'a} et al. (\cite{toala_2017}) suggested that this structure is a stellar bow shock  driven by the HMXB. The HMXB has a well-measured proper motion, which yields a runaway peculiar velocity at the adopted distance. The inferred direction of the peculiar motion deviates by about 60 degrees from the approximate axis of symmetry of the cavity. This difference cannot be attributed to the errors in the peculiar velocity, as it is well constrained. Also, considering the magnitude of the peculiar velocity of the system of about 70 $\mathrm{km}\, \mathrm{s}^{-1}$, the peculiar ambient medium velocity would have to be considerably large to produce such deviation. Another point to consider is the absence of the emission enhancement along the boundary of the cavity. The brightness along the rim of the tentative bow shock is practically the same as in the surrounding cloud to the Galactic southeast. This poses problems for the bow shock interpretation. A possible alternative could be a partial cavity, possibly shaped by the feedback from the system, or a chance alignment of a foreground/background structure.

\subsubsection{EXO 1722-363 nebula}
The system is projected atop a crescent-shaped irregular mid-infrared nebula. Owing to a considerable distance to the system, there is no proper motion information available, so it is not possible to investigate whether there exists a connection with the orientation of the nebula and the systemic peculiar motion. The nebula is not evenly bright; as can be seen in Fig.~\ref{EXO_1722-363_nebula}, the east part of the nebula is significantly brighter. The brightening could be related to the systemic motion, however, it is impossible to ascertain given the lack of data. The system may be passing through a larger ISM cloud and shock only a part of it, while heating some of the unshocked material as well, which could explain the fainter filament that is projected downstream of the tentative bow shock. The nebula could be interpreted as a partial wind-blown bubble, however, the system is lying significantly off center. Another possibility is that the filament projected onto the system is a part of an interstellar cloud crossing behind or in front of the HMXB, passing closest to the system on its east-projected side. We classify this structure as ambiguous.

\subsubsection{XTE J1739-302 filament}
The system is adjacent to a fine mid-infared arc or filament, situated to the Galactic east, and its approximate axis of symmetry is oriented in the same direction. The proper motion data for the system yield a runaway peculiar velocity in the direction approximately to the Galactic northeast, albeit with considerable errors (see Table~\ref{HMXB_gaia}). For this reason, it is not meaningful to investigate the alignment of the peculiar velocity vector with the symmetry axis of the arc. Adding to this, the arcuate filament is not significantly brighter than the surrounding emission to the Galactic south that it seems to be linked with, as would be expected for the heated dust piled at the bow shock front. This implies that the filament cannot be interpreted as a stellar bow shock, and is, most likely, not related to the HMXB system.

\subsubsection{AX J1841.0-0536 cavity}
The system is projected into a cavity of a reduced 24 $\mathrm{\mu}$m emission. Taking the peculiar velocity at face value, the system moves rather slowly for a supergiant-hosting HMXB. However, because of its uncertain distance, its peculiar velocity is ill-determined. Interestingly, the system seems to be a part of a small star group, possibly making up a star cluster. The distance to the cavity is not known, as it is not catalogued. However, the presence of a HMXB together with the surrounding stars in its center, make it possible that the cavity is shaped by stellar feedback.  

\subsubsection{XTE J1855-026 structure}
The system seems to be located in a region of reduced 24~$\mathrm{\mu}$m emission, bracketed by regions of stronger emission to the Galactic south and northwest, possibly making up a partial bubble. The system is a likely runaway owing to its position away from the Galactic plane. The precise kinematics of the system is hard to constrain because of the errors in the parallax measurement, however, none of the bright rims on the either side of the HMXB can be interpreted as stellar bow shocks.

\subsection{Paucity of bow shocks around HMXBs}

Our search for bow shocks driven by HMXBs yielded only one questionable new detection. Even if we consider the structure around \object{GX 304-01} as a bona fide stellar bow shock, together with the bow shocks associated with \object{Vela X-1} and \object{4U 1907+09}, there are only three known HMXB bow shocks (only about 2\% of the HMXBs drive detectable bow shocks). This is considerably lower than the bow shock occurrence around OB runaway stars, although this number has been fluctuating  over the years (about 13\%, van Buren et al.  \cite{vanburen_1995}; $\sim$40\%,  Huthoff \& Kaper \cite{huthoff_2002}; $\sim$10\%, Peri et al. \cite{peri_2012}; and $\sim$6\%, Peri et al. \cite{peri_2015}). This discrepancy can be explained by the hypothesis proposed by Huthoff \& Kaper (\cite{huthoff_2002}) suggesting different ejection scenarios for HMXBs and OB runaways. OB runaways have predominantly escaped their parent associations or clusters at a relatively early stage, when the cluster was dense and the probability of close encounters and ejections of stars was high. On the other hand, a lot of HMXBs became runaways only after the occurrence of a supernova within the system. Thus, the kinematical age and the distance to their parent clusters should be lower for HMXBs. Therefore, HMXBs are more likely to be still enclosed in the hot and rarefied regions (hot bubbles) that surround OB associations and clusters while the OB runaways have already escaped from these regions. The (isothermal) speed of sound in the ISM increases with temperature, thus, the hotter the medium is, the less likely are runaways moving through it supersonically. Together with the lower ISM density in these regions, this means a smaller chance of bow shock detection. At the time, Huthoff \& Kaper (\cite{huthoff_2002}) were only able to study the difference in bow shock occurrence between the OB runaways and HMXBs on a sample of 11 HMXB systems. The analysis of the clustering between HMXBs and OB associations by Bodaghee et al. (\cite{bodaghee_2012b}) also reinforced this notion, suggesting that the massive binaries that are the progenitors of HMXBs tend to remain gravitationally bound to their birth sites until the supernova explosion in the system. Thus, they acquire their runaway velocity only later on, leaving the association or cluster after it has evolved considerably, evacuating cavities in their surroundings, making the formation of an observable bow shock less likely.

Moreover, in addition to bow shocks associated with stars having large peculiar velocities, the young stars in a star cluster can drive significant outflows, causing the local ISM velocities to deviate from the LSR. Thus bow shocks can also be generated around neighboring stars or stars at cluster outskirts without necessarily having large peculiar velocities (e.g., Povich et al. \cite{povich_2008}). These bow shocks are often coined `in situ bow shocks'and their axes of symmetry point predominantly toward the source of outflows. The effects of outflows abates as the cluster ages and thus by the time the first supernova explosions start to occur in a cluster, possibly giving rise to HMXBs or their progenitiors, the outflows are weaker and the ISM around the cluster is more rarified because of feedback effects. This suggests that the in situ bow shock generation is also less likely for HMXBs as compared to other OB stars. 

All this points to the scenario in which massive binaries, which are progenitors to HMXBs, are not efficiently expelled from star clusters and associations during the early stages of a cluster lifetime. This also implies that the two-step ejection process (Pflamm-Altenburg \& Kroupa \cite{pflamm_2010}) is not at work for a considerable number of HMXBs, making it possible to determine their birth sites via the peculiar velocity measurements.

\section{Summary and conclusions}
We have searched for bow shocks around HMXBs using WISE and MIPS \textit{Spitzer Space Telescope} archival data. Apart from the already known bow shocks (associated with \object{Vela X-1} and \object{4U 1907+09}), we found only one new structure resembling a bow shock, but even this structure cannot be conclusively interpreted as a bona fide bow shock.

The detection of the bow shock candidate around \object{GX 304-01} suggests that this system possesses at least a moderate runaway velocity. The proper motion measurements also seem to support the runaway interpretation of \object{GX 304-01}. The bow shock around \object{GX 304-01}, if real, would be the first bow shock detected associated with a Be/X-ray binary. 

The relative paucity of bow shocks associated with HMXBs as compared to OB runaway stars supports the hypothesis that HMXBs are kinematically younger objects than OB runaways, meaning that most of them are still moving within hot and tenuous medium near their parent cluster or association and thus unable to form an observable bow shock.

We expect that future releases of proper motion and paralax measurements obtained by the mission \textit{Gaia} will considerably improve our knowledge of kinematics of these HMXBs. The planned \textit{James Webb Space Telescope} will also be able to obtain more detailed images of bow shock candidates and nebulosities, allowing us to gain more insight into both the diffuse structures and the HMXBs that are associated with them. 

\begin{acknowledgements}
The author acknowledges support from grant MUNI/A/0789/2017. This research has made use of the NASA/IPAC Infrared Science Archive, which is operated by the Jet Propulsion Laboratory, California Institute of Technology, under contract with the National Aeronautics and Space Administration. This work is based [in part] on observations made with the Spitzer Space Telescope, which is operated by the Jet Propulsion Laboratory, California Institute of Technology under a contract with NASA. This publication makes use of data products from the Wide-field Infrared Survey Explorer, which is a joint project of the University of California, Los Angeles, and the Jet Propulsion Laboratory/California Institute of Technology, funded by the National Aeronautics and Space Administration. This work has made use of data from the European Space Agency (ESA) mission {\it Gaia} (\url{https://www.cosmos.esa.int/gaia}), processed by the {\it Gaia} Data Processing and Analysis Consortium (DPAC; \url{https://www.cosmos.esa.int/web/gaia/dpac/consortium}). Funding for the DPAC has been provided by national institutions, in particular the institutions participating in the {\it Gaia} Multilateral Agreement. This research has made use of the SIMBAD database and the VizieR catalog access tool, both operated at CDS, Strasbourg, France. This publication makes use of data products from the Two Micron All Sky Survey, which is a joint project of the University of Massachusetts and the Infrared Processing and Analysis Center/California Institute of Technology, funded by the National Aeronautics and Space Administration and the National Science Foundation. This paper makes use of data obtained as part of the INT Photometric Hα Survey of the Northern Galactic Plane (IPHAS; www.iphas.org) carried out at the Isaac Newton Telescope (INT). The INT is operated on the island of La Palma by the Isaac Newton Group in the Spanish Observatorio del Roque de los Muchachos of the Instituto de Astrofisica de Canarias. All IPHAS data are processed by the Cambridge Astronomical Survey Unit, at the Institute of Astronomy in Cambridge. The band-merged DR2 catalog was assembled at the Centre for Astrophysics Research, University of Hertfordshire, supported by STFC grant ST/J001333/1. This research made use of Montage. It is funded by the National Science Foundation under Grant Number ACI-1440620, and was previously funded by the National Aeronautics and Space Administration's Earth Science Technology Office, Computation Technologies Project, under Cooperative Agreement Number NCC5-626 between NASA and the California Institute of Technology. This research made use of Astropy, a community-developed core Python package for Astronomy (Astropy Collaboration, 2018).

\end{acknowledgements}


\begin{thebibliography}{}

\bibitem[2017]{altmann_2017} Altmann, M., Roeser, S., Demleitner, et al. 2017, A\&A, 600, L4 

\bibitem[2001]{ankay_2001}Ankay, A., Kaper, L., de Bruijne, J.~H.~J., et al. 2001, A\&A, 370, 170 

\bibitem[2016]{Astraatmadja_2016} Astraatmadja, T. L., \& Bailer-Jones, C. A. L. 2016, ApJ, 833, 119 

\bibitem[2018]{bailer-jones_2018} Bailer-Jones, C.~A.~L., Rybizki, J., Fouesneau, M., et al. 2018, AJ, 156, 58

\bibitem[2001]{bamba_2001}Bamba, A., Yokogawa, J., Ueno, M., et al. 2001, PASJ, 53, 1179

\bibitem[2014]{barentsen_2014}Barentsen, G., Farnhill, H.~J., Drew, J.~E., et al. 2014, MNRAS, 444, 3230 

\bibitem[2010]{benaglia_2010}Benaglia, P., Rib{\'o}, M., Combi, J.~A., et al. 2010, A\&A, 523, A62

\bibitem[2001]{berger_2001}Berger, D. H., \& Gies, D. R. 2001, ApJ, 555, 364

\bibitem[2016]{bland-hawthorn_2016} Bland-Hawthorn, J., \& Gerhard, O. 2016, Annual Review of Astronomy and Astrophysics, 54, 529

\bibitem[2006]{bodaghee_2006}Bodaghee, A., Walter, R., Zurita Heras, J. A., et al. 2006, A\&A, 447, 1027

\bibitem[2012a]{bodaghee_2012a}Bodaghee, A., Rahoui, F., Tomsick, J. A., et al. 2012, ApJ, 751, 113 

\bibitem[2012b]{bodaghee_2012b}Bodaghee, A., Tomsick, J. A., Rodriguez, J., et al. 2012, ApJ, 774, 108

\bibitem[2003]{boroson_2003}Boroson, B., Vrtilek, S.~D., Kallman, T., et al. 2003, ApJ, 592, 516

\bibitem[2008]{chaty_2008}Chaty, S., Rahoui, F., Foellmi, C., et al. 2008, A\&A, 484, 783 

\bibitem[2012]{chen_2012} Chen, C. H., Pecaut, M., Mamajek, E. E., et al. 2012, ApJ, 756, 133 

\bibitem[1998]{chevalier_1998} Chevalier, C., \& Ilovaisky, S.A. 1998, A\&A, 330, 201

\bibitem[2013]{coleiro_2013}Coleiro, A., \& Chaty, S. 2013, ApJ, 764, 185

\bibitem[2013]{coleiro_2013b} Coleiro, A., Chaty, S., Zurita Heras, J.~A., et al. 2013, A\&A, 560, A108 

\bibitem[1986]{corbet_1986}Corbet, R. H. D. 1986, MNRAS, 220, 1047

\bibitem[1999]{corbet_1999} Corbet, R.~H.~D., Marshall, F.~E., Peele, A.~G., et al. 1999, ApJ, 517, 956 

\bibitem[2005]{corbet_2005}Corbet, R.H.D., Markwardt, C.B., \& Swank, J.H. 2005, ApJ, 633, 377

\bibitem[2005]{corbet_2005b} Corbet, R.~H.~D., \& Remillard, R. 2005, The Astronomer's Telegram, 377 

\bibitem[2010]{corbet_2010}Corbet, R. H. D., Krimm, H. A., Barthelmy, S. D., et al. 2010, The Astronomer's Telegram, 2570

\bibitem[2013]{corbet_2013}Corbet, R. H. D., \& Krimm, H. A. 2013, ApJ, 778, 45

\bibitem[2015]{deschamps_2015}Deschamps, R., Braun, K., Jorissen, A., et al. 2015, A\&A, 577, A55 

\bibitem[2010]{drave_2010}Drave, S.~P., Clark, D.~J., Bird, A.~J., et al. 2010, MNRAS, 409, 1220 

\bibitem[2004]{fazio_2004}Fazio, G. G., Hora, J. L., Allen, L. E., et al. 2004, ApJS, 154, 10  

\bibitem[1978]{forman_1978} Forman, W., Jones, C., Cominsky, L., et al. 1978, ApJS, 38, 357

\bibitem[2016]{Gaia16} Gaia Collaboration, Brown, A. G. A., Vallenari, A., et al. 2016, A\&A, 595, A2

\bibitem[2018]{gaia_gdr2_summary} Gaia Collaboration, Brown, A.~G.~A., Vallenari, A., et al. 2018, ArXiv e-prints , arXiv:1804.09365.

\bibitem[2004]{grebenev_2004} Grebenev, S.~A., Ubertini, P., Chenevez, J., et al.  2004, The Astronomer's Telegram, 350 

\bibitem[2015]{gonzales-galan_2015}Gonz\'{a}lez-Gal\'{a}n, A. 2015, ArXiv e-prints 1503.01087

\bibitem[2010]{griffin_2010} Griffin, M.J., Abergel, A., Abreu, A. et al. 2010, A\&A, 518, L3

\bibitem[2008]{gvaramadze_2008}Gvaramadze, V.V., \& Bomans, D.J. 2008, A\&A, 490, 1071

\bibitem[2010]{gvaramadze_2010}Gvaramadze, V.V., Kroupa, P., \& Pflamm-Altenburg, J. 2010, A\&A, 519, 33

\bibitem[2011a]{gvaramadze_2011a}Gvaramadze, V.V., Kniazev, A.Y., Kroupa, P., et al. 2011, A\&A, 535, A29

\bibitem[2011b]{gvaramadze_2011smc}Gvaramadze, V.V., Pflamm-Altenburg, J., \& Kroupa, P. 2011, A\&A, 525, 17

\bibitem[2011c]{gvaramadze_2011b}Gvaramadze, V.V., Roser, S., Scholz, R.-D., et al. 2011, A\&A, 529, A14

\bibitem[2004]{halpern_2004}Halpern, J. P., Gotthelf, E. V., Helfand, D. J., et al. 2004, The Astronomer’s Telegram, 289, 1

\bibitem[1982]{hoffleit_1982}Hoffleit, D., \& Jaschek, C. 1982, Bright Star Catalogue (Yale)

\bibitem[2002]{huthoff_2002}Huthoff, F., \& Kaper, L. 2002, A\&A, 383, 999

\bibitem[2004]{int_zand_2004} in't Zand, J., \& Heise, J. 2004, The Astronomer's Telegram, 362

\bibitem[2018]{jennings_2018} Jennings, R.~J., Kaplan, D.~L., Chatterjee, S., et al.  2018, ArXiv e-prints , arXiv:1806.06076

\bibitem[1973]{jones_1973}Jones, C., Forman, W., Tananbaum, H., et al. 1973, ApJ, 181, L43

\bibitem[2002]{kalas_2002}Kalas, P., Graham, J.~R., Beckwith, S.~V.~W., et al. 2002, ApJ, 567, 999 

\bibitem[1997]{kaper_1997}Kaper, L., van Loon, J.Th., Augusteijn, T., et al. 1997, ApJ, 475, L37

\bibitem[2004]{kaper_2004}Kaper, L., van der Meer, A., \& Tijani, A.H. 2004, The Environment and Evolution of Double and Multiple Stars, Proceedings of IAU Colloquium 191, 21, 128

\bibitem[2006]{kaper_2006}Kaper, L., van der Meer, A., \& Najarro, F. 2006, A\&A, 457, 595 

\bibitem[2016]{kobulnicky_2016}Kobulnicky, H. A., Chick, W. T., Schurhammer, D. P.. et al. 2016, ApJS, 2, 18 

\bibitem[2006]{levenhagen_leister_2006} Levenhagen, R.~S., \& Leister, N.~V. 2006, MNRAS, 371, 252 

\bibitem[2006]{lewin_vdklis} Lewin, W.H.G., \& van der Klis, M. 2006, Compact Stellar X-ray Sources, Cambridge, UK: Cambridge University Press

\bibitem[1997]{lewin_97} Lewin, W.H.G., van Paradijs, J., van den Heuvel, E.P.J., et al. 1997, X-ray binaries, Cambridge, UK: Cambridge University Press

\bibitem[2016]{lindegren_2016} Lindegren, L., Lammers, U., Bastian, U., et al. 2016, A\&A, 595, A4

\bibitem[2018]{lindegren_2018} Lindegren, L., Hernandez, J., Bombrun, A., et al. 2018, ArXiv e-prints , arXiv:1804.09366

\bibitem[2006]{liu_2008} Liu, Q.Z., van Paradijs, J., \& van den Heuvel, E.P.J. 2006, A\&A, 455, 1165

\bibitem[2018]{luri_2018} Luri, X., Brown, A.~G.~A., Sarro, L.~M., et al. 2018, A\&A, 616, A9

\bibitem[2005]{makovoz_2005}Makovoz, D., \& Khan, I. 2005, in Astronomical Society of the Pacific Conference Series, Vol. 347, Astronomical Data Analysis Software and Systems XIV, ed. P. Shopbell, M. Britton, \& R. Ebert, 81

\bibitem[2011]{manousakis_2011}Manousakis, A., \& Walter, R. 2011, A\&A, 526, A62

\bibitem[2009]{masetti_2009} Masetti, N., Parisi, P., Palazzi, E., et al. 2009, A\&A, 495, 121

\bibitem[1978]{mason_1978}Mason, K.O., Murdin, P.G., Parkes, G.E., et al. 1978, MNRAS, 184, 45P

\bibitem[2009]{mason_2009}Mason, A.B., Clark, J.S., Norton, A.J., et al. 2009, A\&A, 505, 281

\bibitem[2010]{mason_2010}Mason, A.B., Norton, A.J., Clark, J.S., et al. 2010,  A\&A, 509, A79

\bibitem[2016]{mayer_2016}Mayer, A., Deschamps, R., \& Jorissen, A. 2016, A\&A, 587, A30

\bibitem[1971]{mcclintock_1971}McClintock, J.E., Ricker, G.R., \& Lewin, W.H.G. 1971, ApJ, 166, L73

\bibitem[2009]{megier_2009}Megier, A., Strobel, A., Galazutdinov, G. A., et al. 2009, A\&A, 507, 833 

\bibitem[2016]{meyer_2016}Meyer, D. M.-A., van Marle, A.-J., Kuiper, R. et al. 2016, MNRAS, 459, 1146

\bibitem[2017]{meyer_2017}Meyer, D. M.-A., Mignone, A., Kuiper, R. et al. 2016, MNRAS, 464, 3229

\bibitem[2006a]{negueruela_2006a}Negueruela, I., Smith, D.~M., Harrison, T.~E., et al. 2006, ApJ, 638, 982

\bibitem[2006b]{negueruela_2006b}Negueruela, I., Smith, D. M., Reig, P., et al. 2006, in ESA Special Publication, ed. A. Wilson, 604, 165

\bibitem[2007]{negueruela_2007} Negueruela, I., Torrej{\'o}n, J.~M., \& McBride, V. 2007, The Astronomer's Telegram, 1239 

\bibitem[2008]{negueruela_2008}Negueruela, I., Casares, J., Verrecchia, F., et al. 2008, The Astronomer's Telegram, 1876

\bibitem[2008]{nespoli_2008}Nespoli, E., Fabregat, J., \& Mennickent, R. E. 2008, A\&A, 486, 911

\bibitem[2014a]{ochsendorf_2014a}Ochsendorf, B.B., Cox, N.L.J., Krijt, S., et al. 2014, A\&A, 563, A65

\bibitem[2014b]{ochsendorf_2014b}Ochsendorf, B.B., Verdolini, S., Cox, N.L.J., et al. 2014, A\&A, 566, A75

\bibitem[2005]{parker_2005}Parker, Q.A., Phillipps, S., Pierce, M.J., et al. 2005, MNRAS, 362, 689

\bibitem[1980]{parkes_1980}Parkes, G.E., Murdin, P.G., \& Mason, K.O. 1980, MNRAS, 190, 537

\bibitem[2012]{peri_2012}Peri, C.S., Benaglia, P., Brookes, D.P., et al. 2012, A\&A, 538, A108

\bibitem[2015]{peri_2015}Peri, C. S., Benaglia, P., \& Isequilla, N. L. 2015, A\&A, 578, A45

\bibitem[1997]{perryman_1997}Perryman, M. A. C., Lindegren, L., Kovalevsky, J., et al. 1997, A\&A, 323, L49

\bibitem[2010]{pflamm_2010} Pflamm-Altenburg, J., \& Kroupa, P. 2010, MNRAS, 404, 1564 

\bibitem[2010]{pilbratt_2010}Pilbratt, G.L., Riedinger, J.R., Passvogel, T., et al. 2010, A\&A, 518, L1

\bibitem[2010]{poglitsch_2010}Poglitsch, A., Waelkens, C., Geis, N., et al. 2010, A\&A, 518, L2

\bibitem[1984]{polcaro_1984} Polcaro, V. F., Bazzano, A., La Padula, C., et al. 1984, A\&A, 131, 229

\bibitem[2017]{postnov_2017}Postnov, K., Oskinova, L., \& Torrej\'{o}n, J. M. 2017, MNRAS, 465L, 119P

\bibitem[2008]{povich_2008}Povich, M.~S., Benjamin, R.~A., Whitney, B.~A., et al. 2008, ApJ, 689, 242

\bibitem[1983]{priedhorsky_1983}Priedhorsky, W.C., \& Terrell, J. 1983, ApJ, 273, 709

\bibitem[2008]{rahoui_2008}Rahoui, F., Chaty, S., Lagage, P. O., et al. 2008, A\&A, 484, 801

\bibitem[1999]{reynolds_1999}Reynolds, A.~P., Owens, A., Kaper, L., et al. 1999, A\&A, 349, 873 

\bibitem[2004]{rieke_2004}Rieke, G.H., Young, E.T., Engelbracht, C.W., et al. 2004, ApJS, 154, 25

\bibitem[2011]{romano_2011}Romano, P., Mangano, V., Cusumano, G., et al. 2011, MNRAS, 412, L30

\bibitem[1986]{sato_1986}Sato, N., Nagase, F., Kawai, N., et al. 1986, ApJ, 304, 241

\bibitem[1866]{secchi_1866}Secchi, A. 1866, Astronomische Nachrichten, 68, 6

\bibitem[2014]{servillat_2014} Servillat, M., Coleiro, A., Chaty, S., et al. 2014, ApJ, 797, 114 

\bibitem[1998]{smith_1998}Smith, D.~M., Main, D., Marshall, F., et al. 1998, ApJ, 501, L181

\bibitem[2016]{smith_2016}Smith, M. A., Lopes de Oliveira, R., \& Motch, C. 2016, Adv. Space Res., 58, 782

\bibitem[2017]{smith_2017}Smith, M. A., Lopes de Oliveira, R., \& Motch, C. 2017, MNRAS, 469, 1502 

\bibitem[2014]{sota_2014}Sota, A., Ma{\'{\i}}z Apell{\'a}niz, J., Morrell, N.~I., et al. 2014, ApJS, 211, 10

\bibitem[2001]{sugizaki_2001}Sugizaki, M., Mitsuda, K., Kaneda, H., et al. 2001, ApJS, 134, 77

\bibitem[1990]{takeuchi_1990}Takeuchi, Y., Koyama, K., \& Warwick, R. S. 1990, PASJ, 42, 287

\bibitem[1989]{tawara_1989}Tawara, Y., Yamauchi, S., Awaki, H., et al. 1989, PASJ, 41, 473

\bibitem[2007]{thompson_2007}Thompson, T.W.J., Tomsick, J.A., in 't Zand, J.J.M., et al. 2007, ApJ, 661, 447

\bibitem[2017]{toala_2017} Toal{\'a}, J.~A., Oskinova, L.~M., \& Ignace, R. 2017, ApJ, 838, L19 

\bibitem[2007]{tomsick_2007} Tomsick, J.~A., Chaty, S., Rodriguez, J., et al. 2007, The Astronomer's Telegram, 1231

\bibitem[1988]{vanburen_1988}van Buren, D., \& McCray, R. 1988, ApJ, 329, L93 

\bibitem[1995]{vanburen_1995}van Buren, D., Noriega-Crespo, A., \& Dgani, R. 1995, AJ, 110, 2914

\bibitem[2000]{heuvel_2000}van den Heuvel, E.P.J., Portegies Zwart, S.F., Bhattacharya, D., et al. 2000, A\&A, 364, 563

\bibitem[1989]{oijen_1989}van Oijen, J.G.J. 1989, A\&A, 217, 115

\bibitem[2002]{verrecchia_2002} Verrecchia, F., Negueruela, I., Covino, S., et al. 2002, The Astronomer's Telegram, 102

\bibitem[2015]{walter_2015}Walter, R., Lutovinov, A.A., Bozzo, E., et al. 2015, A\&ARv, 23, 2

\bibitem[1988]{warwick_1988}Warwick, R.S., Norton, A.J., Turner, M.J.L., et al. 1988, MNRAS, 232, 551

\bibitem[2004]{werner_2004}Werner, M., Roellig, T., Low, F., et al. 2004, ApJS, 154, 1

\bibitem[2010]{wright_2010}Wright, E. L., Eisenhardt, P. R. M., Mainzer, A. K., et al. 2010, AJ, 140, 1868

\bibitem[2017]{zacharias_2017} Zacharias, N., Finch, C., \& Frouard, J. 2017, AJ, 153, 166 

\bibitem[2002]{ziolkowski_2002}Ziolkowski, J. 2002, Memorie della Societa Astronomica Italiana, 73, 1038

\bibitem[2006]{zuritaheras_2006}Zurita Heras, J.A., De Cesare, G., Walter, R., et al. 2006, A\&A, 448, 261




\end{thebibliography}
\end{document}